\documentclass[sigconf, nonacm, pdfa]{acmart}
\usepackage[a-2b]{pdfx}

\AtBeginDocument{%
  }

\newcommand\vldbyear{2025}
\newcommand\vldbworkshop{The 2nd International Workshop on Data-driven AI
(DATAI)}
\newcommand\vldbauthors{\authors}
\newcommand\vldbtitle{\shorttitle} 
\newcommand\vldbavailabilityurl{https://github.com/dbislab/InQuRe}
\newcommand\vldbpagestyle{empty}

\usepackage{tcolorbox}
\usepackage{tikz}
\usepackage{csvsimple}
\usepackage{natbib}
\usepackage{algorithm}
\usepackage[english]{babel}  
\addto\extrasenglish{  
    
}
\addto\extrasenglish{  
    
}

\usepackage{blindtext}

\usepackage{tcolorbox}

\usepackage{listings}
\usepackage{color}
\definecolor{dkgreen}{rgb}{0,0.6,0}
\definecolor{gray}{rgb}{0.5,0.5,0.5}
\definecolor{mauve}{rgb}{0.58,0,0.82}
\usepackage{subcaption}

\usepackage{pgfplotstable}
\usepackage{pgfplots}
\usepgfplotslibrary{colorbrewer}
\pgfplotsset{compat=1.18}
\pgfplotsset{cycle list/Dark2}
\usetikzlibrary{pgfplots.statistics}
\usetikzlibrary{arrows,shapes,fit,positioning}
\usetikzlibrary{decorations.pathmorphing}
\usetikzlibrary{decorations.markings}
\pgfplotsset{
	only if/.style args={entry of #1 is #2}{
		/pgfplots/boxplot/data filter/.code={
			\edef\tempa{\thisrow{#1}}
			\edef\tempb{#2}
			\ifx\tempa\tempb
			\else
			B
			\fi
		}
	}
}

\begin{document}

\title{The Case for Intent-Based Query Rewriting}


\renewcommand{\shortauthors}{}


\author{Gianna Lisa Nicolai}
\orcid{0009-0008-9424-1548}
\affiliation{%
\institution{RPTU Kaiserslautern-Landau}
 \city{Kaiserslautern}
 \country{Germany}
 \postcode{67663}
}
\email{gianna.nicolai@cs.rptu.de}

\author{Patrick Hansert}
\orcid{0000-0002-7592-3542}
\affiliation{%
\institution{RPTU Kaiserslautern-Landau}
 \city{Kaiserslautern}
 \country{Germany}
 \postcode{67663}
}
\email{patrick.hansert@cs.rptu.de}

\author{Sebastian Michel}
\orcid{0000-0002-2238-0185}
\affiliation{%
 \institution{RPTU Kaiserslautern-Landau}
 \city{Kaiserslautern}
 \country{Germany}
 \postcode{67663}
}
\email{sebastian.michel@cs.rptu.de}

\newcommand{\approach}{\textsf{INQURE}\xspace}

\begin{abstract}
With this work, we describe the concept of intent-based query rewriting and present a first viable solution. The aim is to allow rewrites to alter the structure and syntactic outcome of an original query while keeping the obtainable insights intact. This drastically differs from traditional query rewriting, which typically aims to decrease query evaluation time by using strict equivalence rules and optimization heuristics on the query plan.  Rewriting queries to queries that only provide a similar insight but otherwise can be entirely different can remedy inaccessible original data tables due to access control, privacy, or expensive data access regarding monetary cost or remote access. 
In this paper, we put forward \approach, a system designed for INtent-based QUery REwriting. It uses access to a large language model (LLM) for the query understanding and human-like derivation of alternate queries. Around the LLM, \approach employs upfront table filtering  and subsequent candidate rewrite pruning and ranking. We report on the results of an evaluation using a benchmark set of over 900 database table schemas and discuss the pros and cons of alternate approaches regarding runtime and quality of the rewrites of a user study.
\end{abstract}

%


\maketitle

\pagestyle{\vldbpagestyle}
\begingroup\small\noindent\raggedright\textbf{VLDB Workshop Reference Format:}\\
\vldbauthors. \vldbtitle. VLDB \vldbyear\ Workshop: \vldbworkshop.\\ 
\endgroup
\begingroup
\renewcommand\thefootnote{}\footnote{\noindent
This work is licensed under the Creative Commons BY-NC-ND 4.0 International License. Visit \url{https://creativecommons.org/licenses/by-nc-nd/4.0/} to view a copy of this license. For any use beyond those covered by this license, obtain permission by emailing \href{mailto:info@vldb.org}{info@vldb.org}. Copyright is held by the owner/author(s). Publication rights licensed to the VLDB Endowment. \\
\raggedright Proceedings of the VLDB Endowment. 
ISSN 2150-8097. \\
}\addtocounter{footnote}{-1}\endgroup

\ifdefempty{\vldbavailabilityurl}{}{
\vspace{.3cm}
\begingroup\small\noindent\raggedright\textbf{VLDB Workshop Artifact Availability:}\\
The source code, data, and/or other artifacts have been made available at \url{\vldbavailabilityurl}.
\endgroup
}



\section{Introduction}
\label{sec:introduction}

The volume of digital content in various formats has been growing exponentially for decades. Companies, organizations, and public administrations generate, capture, and convert user-generated or application-generated content, such as personal data, transaction records, social platform interactions, usage and system logs, and more. Much of this data is not widely accessible to the public, but is stored in company-owned data lakes or traditional databases. In contrast, other datasets are available on platforms like AWS S3 or traditional websites, either free of charge, for a fee, or are perhaps already available at the user's location. The benefit of intent-based query rewriting kicks in when a query is formulated over tables that are not directly available to the user or not available at all due to missing access privileges. The query given in SQL needs to be rephrased to operate on available data and keep the user intent intact. Unlike work in classical query rewriting, which uses equivalence rules from relational algebra, a core observation is that the query itself does not have to be equivalent, i.e., it need not return the identical set of tuples. It is also different from approximate query processing, where the task is to return an approximate answer at the benefit of fast query execution of the same query, where the similarity of result tuples or closeness of aggregate values again assesses the answer's accuracy. For intent-based query rewriting, consider a user interested in finding real estate properties built on land with high land values to guarantee a safe investment. 
Knowing that there is such data (e.g., \autoref{fig:berlin_example} Table d), a SQL query could look like the one shown in  \autoref{listing:sqlpropertyland}.
\begin{figure}
\centering
\scalebox{0.98}{%
\begin{tikzpicture}
    \node [shape=rectangle,fill=orange!20, align=center](table1) at (1.0,-2.0) {
            {\scriptsize
    \begin{tabular}{l|c}
    \bfseries Rank & \bfseries  DistrictName 
    \csvreader[head to column names]{order.csv}{}
    {\\\oid\ & \dname  }
    \end{tabular}  }   };

    \node [shape=rectangle,fill=green!20, align=center](table2) at (1.0,-4.4) {
            {\scriptsize
    \begin{tabular}{l|c|c}%
    \bfseries OfferId & \bfseries  DistrictName & \bfseries Price
    \csvreader[head to column names]{offer.csv}{}
    {\\\oid\ & \dname & \price}
    \end{tabular} } };

    \node [shape=rectangle,fill=blue!20, align=center](table3) at (5.4,-4.4) {
            {\scriptsize
    \begin{tabular}{l|c|c}%
    \bfseries GroundID & \bfseries  DistrictName & \bfseries Value
    \csvreader[head to column names]{land_data.csv}{}
    {\\\lid\ & \dname & \price}
    \end{tabular} } };

	\node[] at  (1,-0.85) (labela) { (a) living\_quality\_ranking};
	\node[] at  (5.5,-0.85) (labelb) { (b) avg\_income};

	\node[] at  (1,-3.2) (labelc) { (c) real\_estate};
	\node[] at  (5.5,-3.2) (labeld) { (d) groundvalue};

    \node [shape=rectangle,fill=red!20, align=center](table4) at (5.3,-2) {
            {\scriptsize
    \begin{tabular}{l|c|c}%
 \bfseries  DistrictId & \bfseries DistrictName & \bfseries Income
    \csvreader[head to column names]{avg_income.csv}{}
    {\\\did & \dname & \income}
    \end{tabular} } };
\end{tikzpicture}
}
\caption{Four tables with different information about Berlin districts. Intuitively, if not all are accessible by a user, main insights can be drawn from the others. A case for intent-based query rewriting.}
\label{fig:berlin_example}
\end{figure}

\begin{lstlisting}[label=listing:sqlpropertyland,caption=Sample Query]
SELECT  DistrictName 
FROM groundvalue
GROUP BY DistrictName
ORDER BY AVG(value) DESC
\end{lstlisting}

Now, consider the user is working, e.g., with a data lake that does not contain that specific table but does contain Tables a--c).
They also hold information about districts in the same city that is closely related to the original query.
With human knowledge, the query could arguably be rewritten to one involving information on schools, average income, businesses, public transport, and other factors. Syntactically, the query would be entirely different, but the goal the user had in mind could still be met.  LLMs exhibit human-like comprehension in many tasks and, in particular, do  very well 
in understanding SQL queries and database schemas. They also understand factors that influence---to stick to the above example--- real estate prices or living quality, which are a mixture of many factors perhaps stored in different databases. 
Once provided with an input query, the LLM is (ideally) able to derive the intent and find other forms of queries that deliver the same insights to a user while operating on different tables, like the ones shown in \autoref{listing:sqlpropertylandrewritten}.

\begin{lstlisting}[float=tb,label=listing:sqlpropertylandrewritten,caption=Two queries with the same intent as \autoref{listing:sqlpropertyland}]
SELECT dname                        SELECT dname
FROM real_estate                    FROM avg_income 
GROUP BY dname                      ORDER BY income DESC
ORDER BY AVG(price) DESC            
\end{lstlisting}

With a growing number of available datasets (databases) at hand, this appears feasible but requires an in-depth investigation and evaluation of potential approaches.
Of course, any such approach dependy on the quality of the LLM's outputs.
Assuming that it is feasible, finding one or multiple alternate intent-equivalent queries for a given user query that operate on different datasets can be helpful in many scenarios, including:

\textbf{Enabling Query Execution over Inaccessible or Monetarily Expensive Data:}  It allows executing queries originally phrased over unavailable datasets, for instance, because of access control mechanisms induced by restrictive data ownership or other privacy concerns. Also, obtaining  access to datasets behind a potentially expensive paywall could be mitigated. 

\textbf{Querying Vast Dataset Collections:}
With the advent of data lakes, the number of available datasets may exceed what users can manually inspect for applicability with reasonable effort.
Intent-based query rewriting allows users to formulate queries over imaginary schemas ideal for their information need.
The system then automatically rewrites these using existing datasets.
Hence, users can forgo the tedious task of manually sighting available datasets.

\textbf{Detection of Divergent Facts:}  With multiple different queries that share the same intent, users or automated processes can detect divergent observations and, therefore, potentially uncover untruthful claims or wrong facts derived from false or incomplete data.

\textbf{Faster Query Execution:}
Finding alternate queries, very much like in the traditional sense of query rewriting, can lead to faster query execution if the database tables referenced in the alternate query are smaller, more suitably indexed, or already locally available. However, this assumes that the time needed to rewrite a query does not overshadow the positive effect on the query runtime. For remote access to LLMs, as we do in this paper, this seems infeasible. 

\subsection{Problem Statement}

Given a SQL query $q$ and a set of tables $T$ by their schema information, we want to determine a ranked list of 
query rewrites $R$ where each $r_i\in R$ gives the same insights as $q$ as intended by the user and is executable over a subset of the tables in $T$. If tables come from different database schemata, we assume they carry a prefix that uniquely identifies the table in $T$.

\subsection{Contributions and Outline}

With this work, we make the following novel contributions:

\begin{itemize}
\item We introduce the problem of intent-based query rewriting.
\item With \approach, we present a first working solution to this problem.
\item We describe the workflow details, including alternate solutions for parts of the building blocks.
\item We discuss problems arising from the use of an LLM for query rewriting and show how we circumvented their hallucinations.
\item We report on the findings of an experimental evaluation that assesses the meaningfulness of rewrites by
means of a user study and further report on metrics like runtime and LLM usage cost.
\end{itemize}

The remainder of this paper is organized as follows: \autoref{sec:relatedwork} reviews related work. \autoref{sec:approach} presents a high-level view of our approach and motivation for the individual building blocks that are then discussed in \autoref{sec:phase1}  and \autoref{sec:phase2}.  \autoref{sec:experiments} reports on the setup and findings of our experimental evaluation, while \autoref{sec:conclusion} concludes the paper and gives an outlook on future work.


\section{Related Work}
\label{sec:relatedwork}

\paragraph{Query Rewriting}

The generic concept of query rewriting~\cite{DBLP:conf/sigmod/AikenWH92,dbcompletebook,DBLP:conf/vldb/AhmedLWDSZC06} is a well-studied key ingredient of relational database systems. An SQL query submitted to the database engine is transformed into an internal tree-structure representation typically referred to as a query plan. Query rewriting then describes the task of modifying the query to a query that produces the same output but can be executed faster or with less consumed resources, depending on the objective. The classic textbook example of query rewriting transforms the query plan using equivalence rules over relational algebra operators and a set of heuristics for improving plans, like pushing down selection operators to eliminate tuples that will not be part of the result as early as possible. Here, equivalence means that the optimized plan still delivers the exact same result tuples as its unmodified version. More advanced techniques consider the non-trivial task of query unnesting~\cite{DBLP:journals/tods/Kim82,DBLP:conf/icde/SeshadriPL96,DBLP:conf/sigmod/MumickFPR90,DBLP:conf/btw/0001K15,DBLP:conf/btw/000125}, as nested queries with dependencies of inner queries to the outer part of the query can lead to poor performance.  
Ordonez \cite{DBLP:journals/tkde/Ordonez10} focuses on optimizing recursive SQL queries using indexing and relational algebra rules. There are also different approaches for optimizing nested queries, summarized and corrected by Ganski and Wong~\cite{DBLP:conf/sigmod/GanskiW87}.

The mentioned approaches mostly keep the SQL queries equivalent regarding a certain set of rules. However, as mentioned above, this is not generalizable to all kinds of SQL queries. Therefore, Dong et al.~\cite{DBLP:journals/pvldb/DongLZYMW23} use synthesis-based techniques for rewriting queries. Although their system does not need predefined rules, it still does not generalize to complicated queries.  Thus, some more recent approaches try to incorporate machine learning into the process. These methods do not always keep the query equivalent regarding rules as in traditional query optimization. However, they still always maintain the result of the query. One of these approaches is proposed by Zhou et al. \cite{DBLP:journals/pvldb/Zhou0WLSZ23}. Their approach combines traditional rule-based methods with learned methods to overcome the problem of rewriting not being applied adaptively. 

Liu and Mozafari \cite{DBLP:journals/corr/abs-2403-09060} go one step further by using LLMs for query rewriting. They employ natural language rewrite rules to guide the LLM in the process of rewriting the query. Their approach shows that LLMs produce promising results for query rewriting and can adapt to different queries. However, one drawback of their approach is that rewrites from LLMs can be incorrect and thus sometimes need correction. This adds a further step in their rewriting process. Ye et al.~\cite{DBLP:journals/corr/abs-2310-09716} also use LLMs for query rewriting. While their goal is not to optimize the query performance, they still use the LLM as a rewriter. To further improve the LLM's result quality, they propose to use distillation, a technique incorporating fine-tuning of the LLM. Their approach also shows promising results.

The work on SODA by~\citet{DBLP:journals/pvldb/BlunschiJKMS12} proposes ways to expand queries to capture tables referring to identical or similar concepts not referenced in the original query. Querying for a table \textit{clients} would then also include the table \textit{customers} as both names are deemed synonymous by their employed ontology. This does not change the query but makes additional tables that carry identical or similar schemas usable. With our use of LLMs, which have a deep understanding of natural language, concepts like synonymy are directly usable, too. In fact, the prompt we send to the LLM contains explicit hints to use such concepts.

\paragraph{SQL Query Similarly}
Since our goal is not to generate an equivalent query but one with the same intent, we need to evaluate how similar the generated query's intent is to the original one.
Traditional measures for query similarity often work syntactically and may be as simple as applying the Levenshtein distance~\cite{levenshtein1966binary} to the query string.
More complex measures may compare the columns referenced in different parts of the query~\cite{DBLP:journals/tkde/KulLXCKU18,DBLP:journals/dke/BorodinK20} or the distance between the SQL syntax trees~\cite{DBLP:journals/dke/BorodinK20}.
To incorporate the semantic meaning of the queries, \citeauthor{DBLP:journals/corr/abs-2403-14441} have proposed a graph-based approach, which represents queries as nodes in a graph and then measures their distance as the lowest-cost path between them \cite{DBLP:journals/corr/abs-2403-14441}.
For our approach, queries may differ significantly in their structure, as we are not interested in equivalent queries but rather in queries that yield the same intent.
Thus, simple syntactic measures are not applicable.

Other approaches, like \citet{DBLP:journals/dke/BorodinK20}, compare the results of the queries as a proxy for the intent.
Similarly, \citeauthor{DBLP:conf/cikm/GuoCXZ11} derive an intent similarity for document retrieval from snippets of the documents \cite{DBLP:conf/cikm/GuoCXZ11}.
\citet{DBLP:journals/tkde/ArzamasovaBGSS20} derive a similarity measure for clustering queries with similar user interests based on the access areas of the queries.
For our approach, none of these measures is applicable since the results of the original query cannot be retrieved.
Instead, we opted to use embeddings or the LLM itself to derive a similarity measure, as discussed in \autoref{sec:ranking}.


\section{The \approach Workflow}
\label{sec:approach}

\begin{figure}
\begin{tikzpicture}
\node[inner sep=0pt] (x) at (0,0) {   \includegraphics[width=1.0\columnwidth]{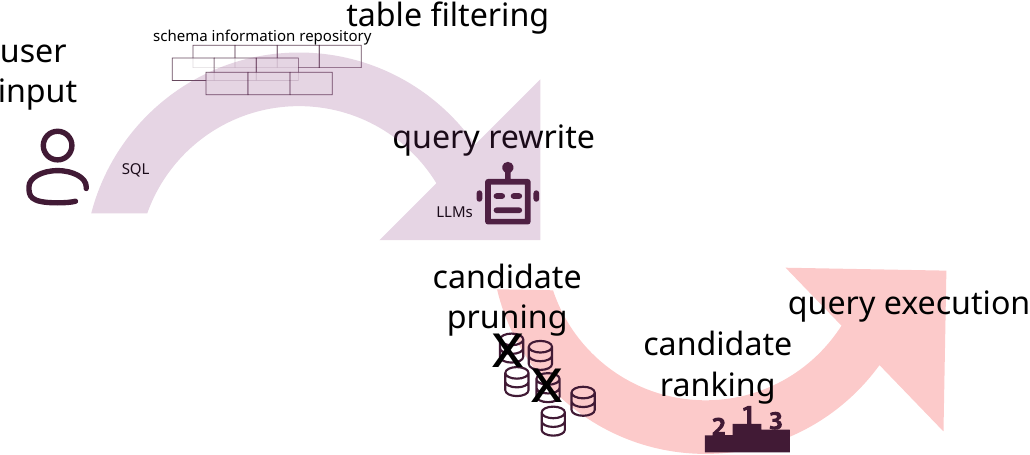}};
\end{tikzpicture}
\caption{Overall workflow consisting of two consecutive phases: (1) Table Filtering and Rewriting and (2) Candidate Cleanup and Execution}
\label{sec:overall}
\end{figure}

\autoref{sec:overall} illustrates the two main phases of \approach. We dive into the details of the phases and sub-phases below but want to give a brief high-level walkthrough and rationale before.
The core element of the workflow is the access to an external LLM that can understand and also produce SQL queries and natural language text. 
Large Language Models (LLMs) like ChatGPT have revolutionized how machines generate text, understand code, and process SQL by leveraging deep learning and vast datasets. They persistently improve and
the latest models show a human-like level of comprehension and even creativity.

While using an LLM can still be feasible given a smaller number of tables, it can cause problems if very many tables are available. 
First, it can result in very high monetary costs (for not free-of-charge LLMs) and high latency due to long requests. Second, and perhaps more limiting, is the rise of LLM ``hallucinations.'' They can occur when there are too many options to choose from, which can result in improper rewrites, and despite the generally amazing performance of ChatGPT to handle SQL, these rewrites can be far-off, cf.~\autoref{listing:hallucination}. While one may argue that moving the attention from baseball to soccer is reasonable as both are sports, querying for all team names is a stark contrast to aiming at the best teams.

\begin{lstlisting}[label=listing:hallucination,caption=Effect of LLM hallucination,float=tb,xleftmargin=3.5em]
--original query               --proposed rewrite
SELECT best_team               SELECT team_long_name
FROM baseball;                 FROM soccer_1_Team;
\end{lstlisting}

To counter this, we employ an upfront table filtering phase to put the focus on seemingly appropriate tables that are then fed into the LLM.
Then, the LLM is asked to provide not only one but multiple rewrites to increase the chance of finding feasible rewrites out of the proposed ones, as it is observed and also reasonable to assume that the first pick by the LLM can be wrong. The proposed rewrites still need to be cleaned up and organized to remove non-executable queries and avoid proposing near-duplicate rewrites to users that differ only marginally, like in a superfluous predicate or a difference in the \texttt{SELECT} clause.


\section{Phase 1: Filtering and Rewrite Generation}
\label{sec:phase1}

The goal of the first phase is to generate a set of candidate rewrites for a given set of tables $T$ and the input SQL query.
As described before, giving too many tables to the rewriter could result in high costs and hallucinations when working with LLMs.
Intuitively, this can be mitigated by reducing the number of available tables to a meaningful subset.
On the other hand, when providing too few or the wrong tables, the rewriter might be unable to produce any relevant rewrites.
For this reason, we first investigate appropriate techniques to prefilter the tables.

\subsection{Table Filtering}
\label{sec:filtering}

We want to eliminate tables unrelated to the topic of the query.
If we only use tables that are, e.g., synonyms, we could miss 
interesting tables that are, e.g., correlated to the original ones. 
We considered multiple approaches for table filtering. One idea was to use knowledge 
databases like WordNet~\cite{DBLP:journals/cacm/Miller95,fellbaum1998wordnet}. 
Such databases include topics or words and their relationships.
Thus, they would be usable to find related topics for the tables in the query. There are two problems with this: 
First, we would still need to somehow identify the topics from the tables or extract the words making up the table names. Generally, extracting topics from a document could be 
done with latent semantic analysis (LSA), as described by \citet{DBLP:journals/jasis/DeerwesterDLFH90}. However, LSA focuses on texts and not on single words like table names 
and can therefore not be used here. The other reason against knowledge databases is that 
LLMs also include much of their knowledge since they are trained on a lot of data. 
Therefore, it is easier to directly use an LLM to deduce topics from the 
tables and compare those using the knowledge from its training---which is possibly more 
than the knowledge in such a knowledge database.
Compared to a knowledge database, the only drawback of an LLM might be the cost, as some knowledge databases are openly available while many LLMs are fee-based.

Overall, we investigated the following three different approaches for table filtering:

\paragraph*{\textbf{Embedding Filter:}} We decided to apply word embedding on the query tables and the tables provided in $T$.
However, we cannot directly supply the unaltered table names into an embedder, as they can be a combination of different words. Thus, an embedder designed to work with single words might not know the compound table names. If a word is not known by an embedder, it may end up being represented by the empty vector, which should be avoided as it is an inadequate representation.
To prevent this, we used the \textit{en\_core\_web\_lg} pipeline \cite{spacy/models/encore} from spaCy \cite{honnibal2020spacy} since it also incorporates tokenization and annotation of words. After multiple pipeline steps, the tokenized words can be vectorized and averaged to represent the table. We use the largest available model for the English language, which generally has a higher accuracy, according to spaCy's website \cite{spacy/models}. A table passes the filter if the cosine similarity between its spaCy embedding vector and any vector of the tables referenced in the query is larger or equal to a parameter $\epsilon$.
For our evaluation, we set $\epsilon$ to 0.4.
This approach is pretty simple, so we do not expect it to perform exceptionally well, especially since we need some tables that are only very vaguely connected to pass the filter, e.g., \textit{groundvalue} and \textit{income} from the introductory example.

\paragraph*{\textbf{Simple LLM Filter:}} We wrote an LLM prompt that takes all our database tables and asks which ones are relevant for rewriting the given SQL query. One thing to keep in mind is the limited amount of in- and output tokens one API call of the LLM can handle.  The so-called context window restricts the combination of input, output, and reasoning tokens. For instance, in GPT-4o, the context window currently has a size of 128,000 tokens, while the number of output tokens is further limited to 16,384 tokens at most \cite{openai/models/gpt4}. If we now want to give a prompt including many table names to the LLM, these limits may be exceeded. Therefore, we partition our table names into packages that are small enough for the LLMs context window and output size. The maximum package size can be calculated from the average amount of tokens needed for one table name, the maximum output length, and the expected percent of returned tables, i.e., how many of the tables are deemed relevant.

We use the following prompt to the LLM: 
\textit{``I want you to decide which from the given tables are useful to answer a given query.  
It is important that the chosen tables can help to answer the query either directly or indirectly.  
If a table cannot be used directly, think about if it can be used due to it being correlated regarding human intuition.  
E.g. tables on park locations and crime rates could help to know areas with high rent.  
So, in conclusion, the tables should be usable to answer the query such that the information gain stays the same for a human.  
Synonyms, hyponyms, correlations and similar topics should be important when choosing the usable tables.  
I have the following SQL query:  
\{query\}  
Here are tables from my database (just the names):  
\{name\_list\}  
If none of these tables are usable, then only respond with 'No tables usable'.  
If some tables are usable, only respond with the names, separated by semicolons.''}

\paragraph{\textbf{Complex LLM Filter:}}  Here, we just ask the LLM for some table names it would suggest for rewriting a given query \emph{without} providing it with the tables that can be accessed. We instruct the LLM with a prompt like \textit{``The tables should be usable to answer the query such
that the information gain stays the same for a human.
Synonyms, hyponyms, correlations, and similar topics should be
important when choosing the usable tables.''}, which produces for a
query \texttt{SELECT best\_team FROM baseball;} the result \textit{
baseball\_teams; mlb\_teams;  team\_rankings;
sports\_statistics; baseball\_standings; fan\_ratings; championship\_winners;
sports\_clubs; stadium\_locations; player\_performance;
 sponsorship\_deals}. We then compare the suggested tables to our database tables. A database table is considered relevant if the cosine similarity of its name's embedding to that of at least one suggested table is higher than or equal to a parameter $\gamma$.
 We set $\gamma$ to 0.7 for our evaluation.

\subsection{LLM-based Rewrites}

To rewrite the query using an LLM is the central step of \approach.
For a given input query, we want to generate \(n\) rewrites that are as diverse as possible.
To this end, we investigate the following two alternatives:

\paragraph{\textbf{Simple Rewriting:}} 
This straightforward approach is inspired by \citeauthor{DBLP:journals/corr/abs-2403-09060}'s prompts for query rewriting \cite{DBLP:journals/corr/abs-2403-09060}. The general idea here is to describe to the LLM what exactly it should do.
To avoid non-executable aggregates of columns, like averages on string columns, we provide the data types along with the table and column names.
Moreover, the LLM is given all foreign keys that a table might have. We do this to improve the joining behavior in the suggested rewrites. Using all this information we prompt the LLM to produce a certain number of diverse SQL rewrites.

We use the following prompt to the LLM:  \textit{``I have the following SQL query:  
\{query\}  
I do not have access to the tables needed in the query.  
I do have the following tables in my database (written in the format:
table: column1 type1, column2 type2 $\backslash$n Foreign keys (if existent)):  
\{filtered\_tables\}  
Keep the foreign keys in mind if you join any tables.  
I want to have queries that keep the same human intent and satisfy the information need as the given query.  
These new queries should use the provided tables and columns from my database.    
Give me \{n\} alternative queries. They should be as diverse as possible.  
Only give the SQL queries and nothing more. Use a semicolon to separate the queries.''
}

After receiving the answer to our prompt, we can separate the produced queries and start with the post-processing phase if enough rewrites are found.

\paragraph{\textbf{Rewriting via Natural Language:}} 
Inspired by the encouraging results of using LLMs in NL2SQL, we do not give an SQL query to the LLM but a natural language (NL) text. That means, before prompting the LLM for the rewrite, we use another LLM request to get the query's intent in textual form. Along with it, we then include the tables with their foreign keys and the columns with their types in the prompt, just like in the simple approach. We also include the number of queries we want to produce. The complete prompt can then be posed to the LLM.
The answer from the LLM to the prompt contains the alternative queries in the same format as in the simple approach. This allows us to process the reply in the same way from here on out.

Compared to the simple approach, we now need one more API call and, thus, more tokens and time. However, since the intent is already abstracted from the SQL query, the LLM can potentially write more general queries without adhering too much to the original query. Since a query with the same intent does not need to be structurally or semantically similar to the original query, this might give this approach an advantage over the simple approach.


\section{Phase 2: Pruning and Ranking}
\label{sec:phase2}

The goal of the second phase of \approach is to take the candidate rewrites from the first phase and return them to the user in a usable manner.
To this end, we first prune candidates with joins on non-joinable columns.
We then rank the remaining candidates based on how well they match the intent of the original query.
Finally, we correct non-executable candidates where necessary.

\subsection{Rewrite Pruning}
\label{sec:pruning}

Here, we want to eliminate queries that will not work with our schema. An example of such a query can be found in \autoref{fig:pruned-query-example}. This query may seem reasonable for finding students with pets. However, with more knowledge of the database schema, we know that the tables for dorm students and students with pets do not originate from the same database. Thus, the IDs given to the students in both of these tables will not match with each other, resulting in false query results. Therefore, we want to eliminate such queries.

\begin{lstlisting}[language=sql, label=fig:pruned-query-example,caption={Example of a Prunable Query},float=tb]
SELECT CONCAT(Fname, ' ', LName) AS full_student_name
FROM dorm_1_Student d, pets_1_Has_Pet hp, pets_1_Pets p 
WHERE d.StuID = hp.StuID AND hp.PetID = p.PetID 
       AND p.PetType = 'Cat'
\end{lstlisting}

\subsection{Rewrite Ranking}
\label{sec:ranking}

After pruning, we want to rank the remaining queries such that better-fitting ones appear earlier in a similar manner to web search.
To this end, we consider two different similarity measures: intent-based similarity and structural similarity.
Using these measures, we rank the rewritten queries either directly or using a Maximal Marginal Relevance (MMR)~\cite{DBLP:conf/sigir/CarbonellG98} approach, as described below.

\subsubsection{Intent-Based Similarity}

For ranking the rewrites, we need to measure how similar the intent of the rewritten query is to the original one.
As discussed in \autoref{sec:relatedwork}, traditional approaches for measuring query similarity are not applicable in our case.
They focus either on syntactic similarity, which is allowed to differ widely in our case, or on the results of the queries, which we cannot retrieve for the original query.
Instead, we investigate two approaches employing embeddings or the LLM itself to derive an intent-based similarity measure.

As our first measure, we compare tables to each other based on their spaCy \cite{honnibal2020spacy} 
embedding using a large model \cite{spacy/models/encore} in an approach we coined \textbf{Embedding Similarity}.
The idea behind it is that different queries with the same intent should still use tables that are related to each other in some way. 
Here, we try and capture this relation with the embeddings
and consider the average pairwise cosine similarity between embeddings of tables of the rewritten query and the original query.
We use our own tokenization for splitting the table names to be sure that the table names are split according to both snake case (\textit{snake\_case}) and camel case (\textit{CamelCase} or \textit{camelCase}), as both often appear in table names.

As an alternative, we investigate \textbf{LLM Similarity} Scores:
We query the LLM with the rewrites and the respective original query. The task of the LLM is to assign a score between $0$ and $1$ to each rewrite that reflects how similar its intent is to the original query. However, LLMs cannot count. Depending on the amount of given queries, not enough similarity scores may be returned by the LLM (i.e., not one score per query). This problem can be solved by giving the rewrites to the LLM in batches. Returning the correct number of scores works most of the time for a small number of queries. The only drawback of this batched approach is the increase in runtime and tokens due to multiple API calls, each containing the instructions again.

We use the following prompt for the LLM: \textit{``
I will give you a single SQL query called original query.
I will also give you multiple other SQL queries called alternative
queries.
For each of the alternative queries, I want to calculate its
similarity to the original query.
These similarities should be floating point values between 0 and 1,
and you should use the following guidelines to appoint them:
The similarity between two queries is solely based on their intent.
If the expected result of the queries gives the same insight to a
human being, their similarity should be 1.
This can include results that are correlated (e.g. rent and crime rate
in a city) or that are virtually the same.
If the given insight is similar, but not the same, the value should be
a floating point number between 0 and 1, depending on how high
the similarity is.
If the query intents do not have anything to do with each other, the
similarity should be 0.
For each given alternative query only return the assigned similarity
value between 0 and 1.
Separate the similarity values using semicolons.
Here is the original query:
\{input\_query\}
Here are the alternative queries:
\{alternative\_queries\}
Give me only the similarity values separated with semicolons.''}

\subsubsection{Ranking Algorithms}

We investigate two options to compute a final ranking of the remaining candidates. First, a \textbf{purely intent-based ranking} that only considers one of the two similarity measures shown above.
However, the LLM may produce multiple rewrites that are just slight syntactic variations of each other.
If these are a good fit according to the similarity measure, other semantically different rewrites may be pushed so far down the ranking that the user never sees them.
Therefore, we augment the intent-based similarity with a structural similarity between candidates of the ranking to boost diversity in the ranked candidates as our second ranking algorithm.
For search engine results, the \textbf{Maximal Marginal Relevance (MMR)}~\cite{DBLP:conf/sigir/CarbonellG98} approach as been proposed to solve this.
It iteratively selects candidate rewrites (or text documents in the original application) in the following manner: First, the rewrite with the highest intent-based similarity is picked. Then, a meta score is calculated greedily for each remaining candidate as its intent-based similarity minus the maximum structural similarity to any of the candidates already picked. The weight $\lambda=0.7$ is used to trade off the influence of intent-similarity and structural dissimilarity.
While \citet{DBLP:journals/tkde/KulLXCKU18} give an overview of different metrics to assess the structural similarity of SQL queries, we opted to use a simple string comparison to assess structural similarity using the inbuilt Python library \texttt{difflib}~\cite{python/difflib}.

\subsection{Query Correction and Execution}

We must check each rewrite to see if it is executable on our database. If not, we want to correct it if possible. This seems relatively easy for syntax errors but gets more complicated if the query, e.g., contains wrong tables. \citet{DBLP:conf/nips/PourrezaR23} describe how an LLM can correct a generated SQL query using a correction prompt. They observed that there is one kind of correction prompt that works best for GPT-4 \cite{openai/models}. We employ a self-correction prompt inspired by theirs. We kept aspects of the original prompt, like the initial instructions and providing the query, the database tables, and the foreign keys. However, instead of providing instructions to correct the query, we provide the LLM with the error message from the database. The correction can proceed in an iterative manner if the proposed corrected query is again not executable.


\section{Experimental Evaluation}
\label{sec:experiments}

All experiments were run on a system with an Intel\(^\circledR\) Core\texttrademark{} i5-8265U CPU and 16 GB of RAM.
We used GPT-4o~\cite{openai/models/gpt4} with the default API parameters in all evaluation runs, the exact version being \textit{gpt-4o-2024-08-06}.
At the time of writing, this was the most performant model from OpenAI~\cite{openai}.
While we did some testing with the cheaper GPT-4o mini as well, it did not perform as well as GPT-4o, e.g., with joins. Hence, we opted to use GPT-4o only.
For \approach, we configured the number of requested rewrites to \(n=5\).

We use the tables from the Spider benchmark~\cite{DBLP:journals/corr/abs-1809-08887}, downloaded from~\cite{spider/dataset}. It contains 206 SQL files, each containing one database schema.
In total, there are 957 tables from 138 different domains. As there are duplicate table names,  we add a prefix to each table and treat all 957 tables as the basis for our algorithms to rewrite the input queries.
We devised a total of ten queries of varying difficulty.
As examples, we show one easy, medium, and hard query in \autoref{listingsamplequeries} and explain the difficulties in rewriting them.

\begin{lstlisting}[escapeinside={<*}{*>},float=t,language=sql,caption=Three sample queries{,} one for each  difficulty level,label=listingsamplequeries]
--Query 2 (easy)
SELECT count(*) FROM college_students;<*\vspace{.78em}*>
--Query 5 (medium)
SELECT school_name, total_budget,
       budgeted_money, invested_money
FROM school_finance
WHERE year = '2024';<*\vspace{.78em}*>
--Query 8 (hard)
SELECT a.author_id, a.author_name,
        COUNT(b.book_id) as total_sales
FROM authors a
     JOIN book_orders b ON a.author_id = b.book_author
WHERE a.author_name LIKE 'M%'
      AND b.purchase_date > '2020-01-01'
GROUP BY a.author_id, a.author_name
ORDER BY total_sales DESC;
\end{lstlisting}

For rewriting Query 2, the difficulty is finding the relevant tables, as there are many tables related to students in the database. Moreover, the aggregation (i.e., count) must still be correct.
For Query 5, there is no table in the database which directly supplies the queried budget.
The rewriter needs to find suitable substitutes, calculate the values from their contents, and then apply the filter condition correctly.
Clearly, this requires a deeper understanding of both SQL and the meaning of the mentioned column names.
Query 8 has more attributes in its select clause and also includes an aggregation. The rewriter needs to find feasible tables and join them correctly, which is more complicated than before, as there are more tables to join. Moreover, the filter condition must be retained, and the right columns must be grouped to ensure the correct accumulation. This is more complex than the previous queries.

As outlined in \autoref{sec:approach}, the individual parts of the framework can be instantiated with different alternatives and combined in configurations, as the different phases are orthogonal. For the rewriting, we evaluate the spaCy Embedding Filter (\textbf{E}), the Simple LLM Filter (\textbf{SLLM}), and the Complex LLM Filter (\textbf{CLLM}), each combined with either Simple Rewriting (\textbf{S}) or NL Rewriting (\textbf{NL}).
For the ranking, we evaluate Intent-based Ranking (\textbf{I}) and \textbf{MMR}, each combined with either Embedding Similarity (\textbf{ES}) or LLM Similarity (\textbf{LLMS}).

To assess the viability of the various combinations, we report on the following primary measures of interest:

\paragraph*{\textbf{Quality:}} To evaluate if a rewrite is feasible, we set up a user study where candidate rewrites for each input query are evaluated by three human evaluators, who assess if the query keeps the same intent as the original query.
The study contained candidates that the configurations under evaluation did not produce.
We consider a rewrite correct if two out of three evaluators marked it as such.

\paragraph*{\textbf{LLM Cost:}} We measure the input and output tokens separately to compare the cost of the API calls. The cost is calculated based on the cost of GPT-4o\footnote{https://openai.com/api/pricing/} as of February 14, 2025:  \$2.50  per 1M tokens input and 
\$10.00  per 1M tokens output.

\paragraph*{\textbf{Runtime:}} Here, we report the wall clock time. As we do not execute the query or any of the rewrites over actual data tables, this only includes the rewriting itself or the cost of the individual components, respectively.

\subsection{Quality of Rewriting}
\label{ssec:quality-rewriting}


\pgfplotstableread[col sep = comma]{plots/study-rewriting-precisions-approaches.csv}{\readdatastudyrewriteapp}

\begin{figure}
			\begin{tikzpicture}
				\begin{axis}[
					ybar,
					width=\columnwidth,
					height=0.5\columnwidth,
					bar width= 0.4em,
					ymin=0,
					ylabel={Precision},
					xlabel={Rewriting Approach},
					xtick= data,
					xticklabels={E+S,E+NL,SLLM+S,CLLM+S,SLLM+NL,CLLM+NL
					},
					xticklabel style={rotate=90},
					major x tick style={opacity=0},
					minor x tick num=1,
					xminorgrids,
					legend pos=north west,
					legend style={draw=none, fill=none},
					]
					\addplot[ybar,draw=none,fill=Dark2-A] table[x=Approach,y=PrecisionFor50General] {\readdatastudyrewriteapp};
					\addplot[ybar,draw=none,fill=Dark2-B] table[x=Approach,y=PrecisionFor50All] {\readdatastudyrewriteapp};
					\legend{Raw Rewrites, Final Rewrites}
				\end{axis}
			\end{tikzpicture}
		\caption{Precision per Rewriting Approach}
		\label{fig:study-rewriting-approach-prec}
\end{figure}
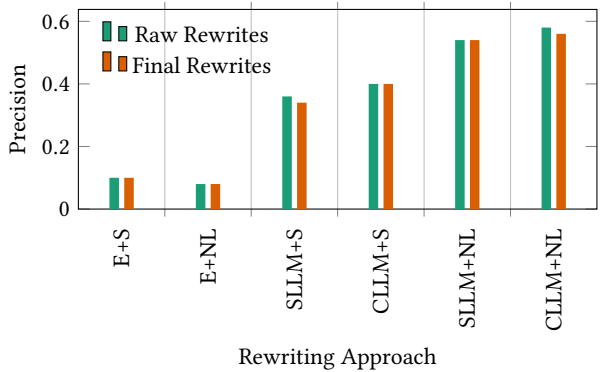

For the rewriting quality, we can both look at the produced rewrites from the first phase and at the rankings from the second phase.

In \autoref{fig:study-rewriting-approach-prec}, we can see the precision of the approach combinations, i.e., how many of the expected total 50 rewrites for all queries were deemed to have the same intent as the respective input query in the user study.
Where fewer than expected rewrites were returned, we count the missing ones as incorrect.
We further distinguish between Raw Rewrites before pruning and correction and Final Rewrites after these steps.
The approaches using the embedding filter each only produced a total of 5 rewrites for simple queries because the filter was too simple to find relevant tables.
As a result, both rewriters perform poorly with this table filter.
This underlines the need to use LLMs for table filtering.

Among the other configurations, those using the NL rewriter perform better than the same approaches with the simple rewriter.
This already answers the question of the better rewriter, leaving us with the task of finding the best-performing table filter.
Overall, both the SLLM and the CLLM filter perform similarly well, with the CLLM filter being slightly better.
It selects more candidate tables.
Besides, configurations with SLLM failed to produce the full number of rewrites for Query~4.
In total, the NL rewriter works well with either filter when considering the final rewrites.


\pgfplotstableread[col sep = comma]{plots/study-rewriting-precisions-queries.csv}{\readdatastudyrewriteQuery}

\begin{figure}
			\begin{tikzpicture}
				\begin{axis}[
					ybar,
					width=\columnwidth,
					height=0.5\columnwidth,
					bar width= 0.25em,
					ymin=0,
					ylabel={Precision},
					xlabel={Input Query},
					xtick= data,
					xticklabels from table={\readdatastudyrewriteQuery}{Query},
					xminorgrids,
					major x tick style={opacity=0},
					minor x tick num=1,
					legend pos=north east,
					legend style={draw=none, fill=none},
					]
					\addplot[ybar,draw=none,fill=Dark2-A] table[x=Query,y=PrecisionGeneral] {\readdatastudyrewriteQuery};
					\addplot[ybar,draw=none,fill=Dark2-B] table[x=Query,y=PrecisionAll] {\readdatastudyrewriteQuery};
					\legend{Raw Rewrites, Final Rewrites}
				\end{axis}
			\end{tikzpicture}
		\caption{Precision per Input Query}
		\label{fig:study-rewriting-query-prec}
\end{figure}
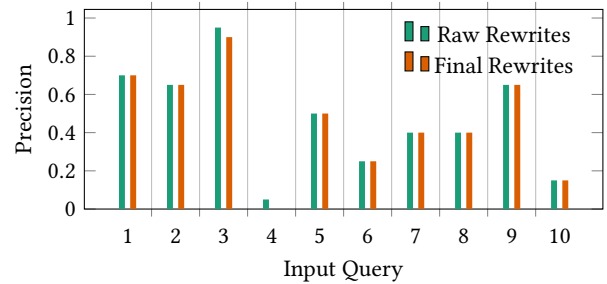

Besides analyzing the precision for each approach, we also look at the queries individually, i.e., how many of the rewrites found for each query by any approach were correct.
Based on the previous observations, we limit this to LLM-based filters.
As before, we count missing rewrites as incorrect.
The resulting precisions can be found in Figure \ref{fig:study-rewriting-query-prec}.
Using them, we can identify the difficulties that need to be solved in intent-based query rewriting by looking at the queries resulting in low precisions.

As can be seen, the hardest ones were Queries 4, 6, and 10.
For Query~4, which asked for students who own a pet cat, the filters struggled to find tables spanning both topics.
With mostly non-joinable tables to work with, the rewriters hallucinated many non-correctable rewrites.
Query~6 was about finding books with a certain length in the order of their customer reviews.
The failed attempts could either again stem from the table filters failing to identify tables representing the customer reviews or from the rewriters failing to interpret the intent of the query correctly.
We also see cases where the rewriters dropped parts of the query, e.g., book's length.
Query~10 asked for drama workshop groups ranked by their popularity per region.
Besides struggling to work with window functions, \approach again struggled to find a good surrogate for the popularity.

We have seen similar issues occur in other queries as well, albeit to a lesser extent.
Therefore, we overall identify a need for table filters with higher recall of relevant tables as an important area for future work.
For the rewriters, we see a need to investigate methods that have an improved understanding of SQL, can better connect the concepts behind data columns, and ensure that all aspects of the input query are preserved in the results.

Except for the fourth query, pruning and non-correctability did not make a big difference. As explained above, this could be because this query brings two relatively different topics together. So, the executability of the rewrites hardly differs due to the hardness of the rewrites. So, the LLM is good at using tables that actually belong together and writing executable SQL.


\pgfplotscreateplotcyclelist{color list ranking}{
	Dark2-B,Dark2-B,Dark2-A,Dark2-A}

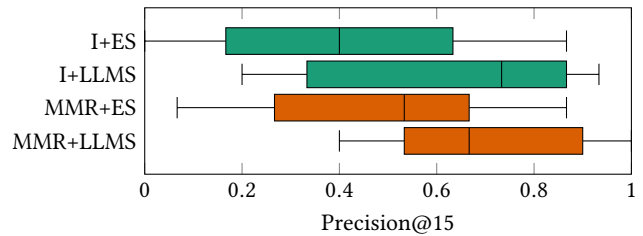
\begin{figure}
	\centering
	\begin{tikzpicture}
		\begin{axis}[
			boxplot/draw direction = x,
			width=.95\columnwidth,
			height=12em,
			ytick= {1,2,3,4},
			yticklabels={MMR+LLMS, MMR+ES, I+LLMS, I+ES},
			major y tick style={opacity=0},
			ymin=0,
			ymax=5,
			xmin=0,
			xmax=1,
			xlabel={Precision@15},
			legend pos=north west,
			cycle list name=color list ranking
			]
		\addplot+[boxplot,draw=black,fill,mark=*,scatter/use mapped color={color list ranking}] table[col sep=comma,y=Precision at 15, only if={entry of Approach is 4}] {plots/study-ranking.csv};
		\addplot+[boxplot,draw=black,fill,mark=*,scatter/use mapped color={color list ranking}] table[col sep=comma,y=Precision at 15, only if={entry of Approach is 3}] {plots/study-ranking.csv};
		\addplot+[boxplot,draw=black,fill,mark=*,scatter/use mapped color={color list ranking}] table[col sep=comma,y=Precision at 15, only if={entry of Approach is 2}] {plots/study-ranking.csv};
		\addplot+[boxplot,draw=black,fill,mark=*,scatter/use mapped color={color list ranking}] table[col sep=comma,y=Precision at 15, only if={entry of Approach is 1}] {plots/study-ranking.csv};
		\end{axis}
	\end{tikzpicture}
	\caption{Precision of the Top-Ranked Results by Approach}
	\label{fig:study-ranking-precision}
\end{figure}

To evaluate the quality of our ranking approaches, we used all rewrites evaluated in the user study and ranked them based on their similarity to the respective original query.
There were, on average, 44.1 rewrites per query.
The boxplot in \autoref{fig:study-ranking-precision} summarizes how many of the top 15 ranking rewrites were correct for each query.
When looking at the results, we can see that the LLM intent similarity is the best in both approaches. This is likely due to the LLM better capturing the real intent and thus the similarity than a simple embedding.

The MMR ranking outperforms the other approach, but just by a small margin. So, even though MMR also looks at the heterogeneity of the queries---which could worsen the precision if there are not many distinct relevant queries---it still performs well in giving out meaningful alternatives as the first 15 queries.

\subsection{Performance and Cost Evaluation}

In this section, we evaluate the performance of our system's components, i.e., the table filters, the rewriter, and the query correction.

\subsubsection{Table Filter}

\begin{figure}[t]
		\begin{tikzpicture}
			\begin{axis}[
				ybar,
				width=\columnwidth,
				bar width= 0.25em,
				ymin=0,
				height=0.5\columnwidth,
				xlabel={Input Queries},
				ylabel={Average Table Count},
				xtick= data,
				xticklabels from table={plots/eval-table-filter-tables.txt}{class},
				major x tick style={opacity=0},
				minor x tick num=1,
				legend pos=north east,
				legend style={draw=none, fill=none},
				legend columns=3]
				\addplot[ybar,draw=none,fill=Dark2-A] table[x=class,y=app1] {plots/eval-table-filter-tables.txt};
				\addplot[ybar,draw=none,fill=Dark2-B] table[x=class,y=app2] {plots/eval-table-filter-tables.txt};
				\addplot[ybar,draw=none,fill=Dark2-C] table[x=class,y=app3] {plots/eval-table-filter-tables.txt};
				\legend{E\,\,,SLLM\,\,,CLLM
}
			\end{axis}
		\end{tikzpicture}
	\caption{Tables found in Average per Query and Filter}
	\label{fig:tables-per-tables-filter}
\end{figure}
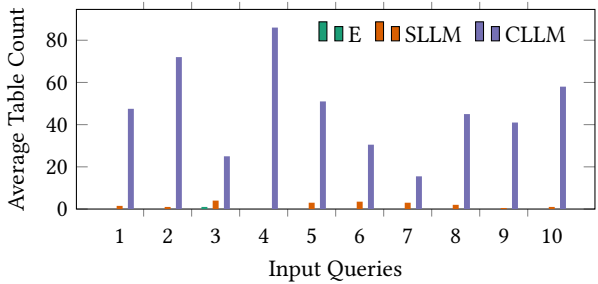

The table filters aim to reduce the number of tables given to the rewriter to a reasonable number of relevant tables.
Although it is hard to quantify the ideal number of tables for a query, we can use the table count to analyze the selectivity of the filters.
Moreover, we can see how much the algorithms generalize from the original tables.
Thus, we will also examine how many of the 957 tables are selected by each filter.
We repeated this twice for each filter and show the average results in \autoref{fig:tables-per-tables-filter}.
The plot is further split by the individual queries since there are stark differences based on the input query.
As we can see, the number of tables deemed relevant not only varies between the queries but also between the approaches.
The CLLM filter finds a lot of tables compared to the other two approaches.
While the spaCy filter often finds zero tables, the SLLM filter finds at least some.
Looking at all approaches, we can conclusively say that spaCy is too restrictive, followed by the SLLM filter, while the CLLM filter is the least restrictive one.

\begin{figure}
	\begin{center}
		\begin{tikzpicture}
			\begin{axis}[
				boxplot/draw direction = x,
				width=\columnwidth,
				height=3cm,
				ytick= {1,2,3},
				yticklabels={CLLM,SLLM,E},
				major y tick style={opacity=0},
				xmin=0,
				xlabel={Runtime [sec]},
				ylabel={Table Filter},
				legend pos=north west,
				]
				\addplot+[boxplot,draw=black,fill=Dark2-C,mark=*] table[y index = 3]{\datatransposed};
				\addplot+[boxplot,draw=black,fill=Dark2-B,mark=*] table[y index = 2]{\datatransposed};
				\addplot+[boxplot,draw=black,fill=Dark2-A,mark=*] table[y index = 1]{\datatransposed};
			\end{axis}
		\end{tikzpicture}
	\end{center}
	\caption{Table Filters Runtimes}
	\label{fig:runtime-table-filter}
\end{figure}
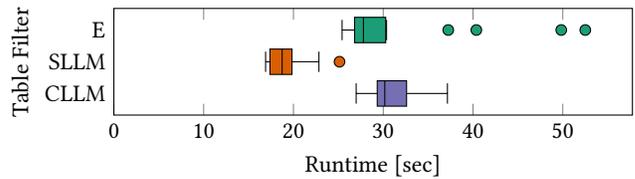

We have also measured the runtime of the above table filter runs.
The results are shown in \autoref{fig:runtime-table-filter}.
As can be seen, the spaCy filter is slower than the SLLM filter despite needing no LLM calls.
Since it is also too restrictive, this clearly shows the value of using LLMs for table filtering.
Between the two LLM-based filters, SLLM is faster than CLLM, as expected.
The latter requires multiple calls to the LLM and, thereby, is slower than the spaCy filter.
However, all three filters take 20 to 30 seconds to run, likely because our benchmark contains close to 1000 tables.
As such, the runtime is preferable to an analyst manually searching relevant tables for minutes, but it still needs much improvement to be usable for faster query execution.

\pgfplotstableread[col sep = comma]{plots/eval-table-filter-runtime.txt}{\readdata}
\pgfplotstabletranspose{\datatransposed}{\readdata}

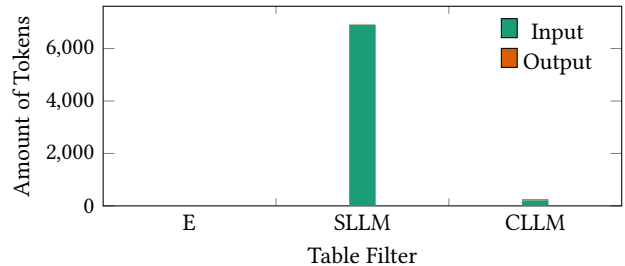
\begin{figure}
	\centering
			\begin{tikzpicture}
				\begin{axis}[
					ybar stacked,
					width=\columnwidth,
					height=0.5\columnwidth,
					ymin=0,
					xmin=0.5,
					xmax=3.5,
					ylabel={Amount of Tokens},
					xlabel={Table Filter},
					xtick= data,
					xticklabels from table={plots/eval-table-filter-tokens.txt}{name},
					major x tick style={opacity=0},
					minor x tick num=1,
					legend pos=north east,
					legend style={draw=none, fill=none},
					legend columns=1]
					\addplot[ybar,draw=none,fill=Dark2-A] table[x=class,y=input] {plots/eval-table-filter-tokens.txt};
					\addplot[ybar,draw=none,fill=Dark2-B] table[x=class,y=output] {plots/eval-table-filter-tokens.txt};
					\legend{Input, Output}
				\end{axis}
			\end{tikzpicture}
		\caption{In- and Output Tokens per Table Filter}
		\label{fig:tokens-table-filter}
\end{figure}

Another interesting measurement is the number of in- and output tokens per execution used by each filter as reported by the API, as these directly correspond to monetary cost.
As before, we ran each filter twice for all queries and considered the average per approach, as seen in Figure \ref{fig:tokens-table-filter}.
There is one filter without any cost, namely the spaCy filter.
It just uses embeddings and no LLM at all.
Hence, there is no token usage connected to this filter.
In contrast, the worst filter regarding tokens and cost is by far the SLLM filter, with nearly 7000 tokens used for input, i.e., 1.7~cents.
This was expected since this filter gives all the existing database tables to the LLM in each call.
In comparison, the CLLM filter only uses around 200 tokens for input, i.e., 0.05~cent.
The number of output tokens for both LLM-based approaches is marginal, with around 15 for SLLM and 50 for CLLM, i.e., 0.01 and 0.05~cents, respectively.
In summary, CLLM is significantly cheaper than SLLM, with only 6\% of the total cost.
However, as discussed before, the runtime of CLLM is 50\% higher than that of SLLM, so the choice of the filter depends on the desired trade-off between cost and runtime.

\subsubsection{Rewriter}
\label{subsubsec:metrics-eval-rewriting-runtime}

We have run both rewriters with \(n=10\) for each query and filter combination that found tables and present the runtime results in the boxplot in \autoref{fig:runtime-rewriting}.
For the simple approach, we can see that the average runtime is a few seconds lower than for the NL-based one. During the execution, the latter makes two calls to the LLM, first getting the query intent and then prompting for the rewrite. Thus, the latency to the API is included twice. In contrast, the simple rewriter only needs one call to the API. This explains the gap between the average runtimes. Whether this is acceptable depends on the user's desired latency of the rewrite system.

For the simple approach, there are also some outliers, which take up to 26 seconds. A reason for this could be the latency of the LLM API as well. If the current load on the API is high, it is slower, and hence, the latency for each single call increases.

\begin{figure}
	\begin{subfigure}[t]{\columnwidth}
		\centering
		\scalebox{.9}{
		\begin{tikzpicture}
			\begin{axis}[
				width=1\columnwidth,
				height=0.5\columnwidth,
				xmin=0,
				ymin=0,
				xlabel={Number of Tables},
				ylabel={Input Tokens},
				separate axis lines,
				ytick= {0,2000,4000,6000,8000,10000},
				legend pos=south east,
				legend style={draw=none, fill=none},
				legend columns=2,
				]
				\addplot+[scatter,only marks,point meta=explicit symbolic,mark=*,scatter/classes={1={color=Dark2-A},2={color=Dark2-B}}] table[y=InTokens,x=NumTables,meta=Rewriting] {plots/eval-rewriting-tokens.txt};
				\legend{S\,\,, NL}
			\end{axis}
			\begin{axis}[
				width=1\columnwidth,
				height=0.5\columnwidth,
				xmin=0,
				ymin=0,
				axis x line*=bottom,
				axis y line*=right,
				xticklabels={},
				ylabel={Cent(\$)},
				major x tick style={opacity=0},
				]
				\addplot+[scatter,only marks,point meta=explicit symbolic,mark=*,scatter/classes={1={color=Dark2-A},2={color=Dark2-B}}] table[y=InCost,x=NumTables,meta=Rewriting] {plots/eval-rewriting-tokens.txt};
			\end{axis}
		\end{tikzpicture}
		}
		\caption{Input Tokens and Cost per Query Rewriting depending on the Amount of Tables found by the Filter}
		\label{fig:tokens-input-rewriting}
	\end{subfigure}
	\hfill
	\begin{subfigure}[t]{\columnwidth}
		\centering
		\scalebox{0.9}{
		\begin{tikzpicture}
						\begin{axis}[
								width=1\columnwidth,
								height=0.5\columnwidth,
								xmin=0,
								ymin=0,
								separate axis lines,
								xlabel={Input Query},
								ylabel={Output Tokens},
								xtick= data,
								legend pos=north west,
								legend style={draw=none, fill=none},
								legend columns=2,
								]
								\addplot+[scatter,only marks,color=Dark2-A,mark=*,point meta=explicit symbolic,scatter/classes={1={color=Dark2-A},2={color=Dark2-B}}] table[y=OutTokens,x=Query,meta=Rewriting] {plots/eval-rewriting-tokens.txt};
								\legend{S\,\,, NL}
							\end{axis}
			\begin{axis}[
				width=1\columnwidth,
				height=0.5\columnwidth,
				xmin=0,
				ymin=0,
				axis x line*=bottom,
				axis y line*=right,
				xticklabels={},
				ylabel={Cent(\$)},
				legend pos=north west,
				legend style={draw=none, fill=none},
				]
				\addplot+[scatter,only marks,color=Dark2-A,mark=*,point meta=explicit symbolic,scatter/classes={1={color=Dark2-A},2={color=Dark2-B}}] table[y=Outcost,x=Query,meta=Rewriting] {plots/eval-rewriting-tokens.txt};
			\end{axis}
		\end{tikzpicture}
		}
		\caption{Output Tokens and Cost per Query Rewriting depending on the Input Query}
		\label{fig:tokens-output-rewriting}
	\end{subfigure}
	\caption{Tokens and Cost of Query Rewriting Approaches}
	\label{fig:tokens-cost-rewriting}
\end{figure}
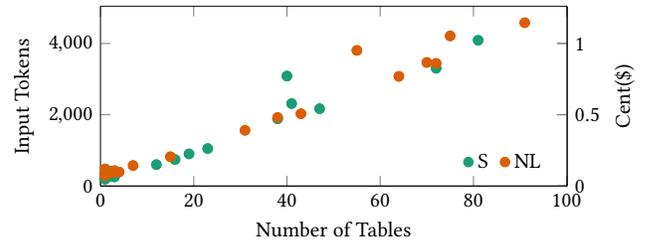
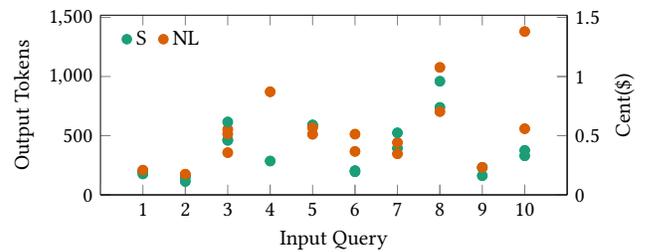

Next, we examine the amount of input tokens used and the resulting costs on the same set of executions as for the runtimes.
As seen in Figure \ref{fig:tokens-input-rewriting}, it primarily depends on the number of tables chosen as relevant by the table filter. This was expected since both approaches incorporate the usable tables in their prompt. Compared to this, the query itself---which is also part of the prompt---hardly affects the amount of input tokens used, especially if there is a large number of tables. Even for the NL approach, which has two requests to the API, no big difference can be seen. As the request to get the intent has only roughly 65 tokens plus the query tokens, its impact is insignificant compared to the table names in the rewrite request.

This is different for the required output tokens, as visualized in \autoref{fig:tokens-output-rewriting}.
Since fewer of them are used in general, the higher amount of output tokens used by the NL approach for the same query is noticeable in most cases.
That aside, the amount of tokens used in the output depends on the individual query.
A reason for this could be the queries differing in complexity and, thus, the length of the rewrites.
For example, Query 2 and its rewrites are much shorter than Query 8.

As the overall monetary cost of LLM calls is directly derived from the number of input and output tokens, it exhibits the same dependencies as described above. Depending on the amount of tables and the query, costs can reach up to 2.5~cents for one iteration of rewriting, i.e., 10 rewrites.
This is, for example, reached by the NL rewriter for Query~10 with 75 input tables. A good table filter is needed to reduce the amount of input tokens and keep the costs low. The output cost can only be reduced by producing fewer rewrites or using the simple rewriter since it is generally a bit less costly. However, as most queries cost below 1 cent, both rewriters are still a feasible solution.


\pgfplotstableread[col sep = comma]{plots/eval-rewriting-runtime.txt}{\readdata}
\pgfplotstabletranspose{\datatransposed}{\readdata}

\begin{figure}
		\begin{tikzpicture}
			\begin{axis}[
				boxplot/draw direction = x,
				width=1\columnwidth,
				height=8em,
				ytick= {1,2},
				yticklabels={NL,S},
				major y tick style={opacity=0},
				xmin=0,
				xlabel= {[sec]},
                xlabel style={at={(axis description cs:1,-0.07)}},
				ylabel={Rewriter},
				legend pos=north west,
				]
				\foreach \n in {2,...,1}{
					\addplot+[boxplot,draw=black,fill,mark=*,scatter/use mapped color={Dark2}] table[y index = \n]
					{\datatransposed};
				}
			\end{axis}
		\end{tikzpicture}
	\caption{Query Rewriter Runtimes}
	\label{fig:runtime-rewriting}
\end{figure}
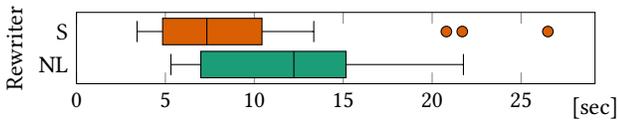

\subsubsection{Ranking}

While we have already investigated the ranking quality in \autoref{ssec:quality-rewriting}, we now also want to see how the approaches perform in terms of quantitative measures.
\autoref{fig:runtime-ranking} shows a boxplot of each ranking approach's runtime for each query.
As expected, MMR is generally slower than its only intent-based counterpart, and the variants using embedding similarity are generally faster than those using the LLM.
This directly presents a trade-off: Approaches that yielded a higher precision in \autoref{fig:study-ranking-precision} are, in turn, slower.
However, we also see that even the slowest approach is about 50\% faster than the table filter or, in other words, about as fast as the rewriter.
As such, all these runtimes are still acceptable but could be improved in future work.

 \pgfplotstableread[col sep = comma]{plots/eval-ranking-runtime.csv}{\readdata}

\begin{figure}
	\begin{center}
		\begin{tikzpicture}
			\begin{axis}[
				boxplot/draw direction = x,
				width=.9\columnwidth,
				height=4cm,
				ytick= {1,2,3,4},
				yticklabels={MMR+LLMS,MMR+ES,I+LLMS,I+ES},
				major y tick style={opacity=0},
				xmin=0,
				ylabel={Ranking Approach},
				xlabel={Runtime [sec]},
				legend pos=north west,
				cycle list name=color list ranking
				]
				\addplot+[boxplot,draw=black,fill=Dark2-B,mark=*] table[y index = 5]{\readdata};
				\addplot+[boxplot,draw=black,fill=Dark2-B,mark=*] table[y index = 4]{\readdata};
				\addplot+[boxplot,draw=black,fill=Dark2-A,mark=*] table[y index = 2]{\readdata};
				\addplot+[boxplot,draw=black,fill=Dark2-A,mark=*] table[y index = 1]{\readdata};
			\end{axis}
		\end{tikzpicture}
	\end{center}
	\caption{Runtimes of Ranking Combinations}
	\label{fig:runtime-ranking}
\end{figure}
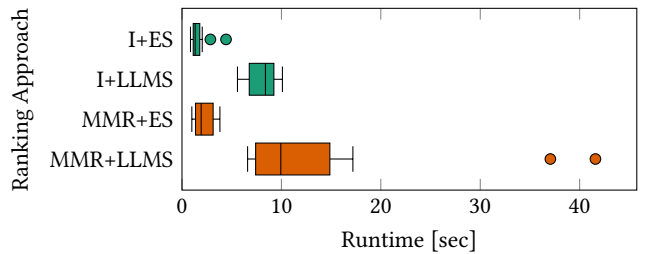

In terms of cost, we note that embedding similarity does not require calls to the LLM and, as such, does not incur any additional API costs.
At the same time, MMR also does not require additional tokens since we compute the structural similarity between queries without any LLM involvement.
Hence, the only costs incurred are the input and output tokens of the LLM similarity.
There, we observe an average total of 0.9~cents, most of which is spent for the average of 3000 input tokens.

\subsubsection{Query Correction}

To see how effective the correction is, we measure the number of iterations needed per non-executable rewrite, i.e., the number of calls to the LLM required to correct the query. Most queries (97 out of 131) are corrected in one iteration.
Only a few outliers need 2 or even 3 iterations (11 and 3).
More iterations were not allowed.
In total, only 20 out of 880 total queries were not correctable, i.e., around 2\%. So, of the 131 queries that needed correction, around 15\% could not be corrected. In 18 of these cases, this was due to the LLM repeatedly using tables that were not in the database. Together with the insight that most queries only needed one correction, we conclude that the correction works efficiently.


\section{Conclusion and Outlook}
\label{sec:conclusion}

We presented \approach, a first working solution to the novel problem of intent-based query rewriting that strives toward query evaluation over inaccessible data by formulating queries over alternate tables that eventually return the same human insights. \approach involves an LLM for 
the core task of rewriting and additional phases to enhance quality and limit cost.
The experimental results showed that it is indeed often possible to find meaningful rewrites and that the monetary cost of involving an LLM is rather negligible. 

It is relatively easy to envision alternate ways to rank or cluster candidate rewrites or to compute structural similarity. Also, tuning the used LLM prompts or ultimately fine-tuning the LLM for this specific task carries the potential to improve the quality of the final rewrites.
We have not yet explored whether intent-based rewrites can also lead to faster query execution. It is clear that the overhead induced by the remote LLM can be prohibitively large for some low-latency queries.
Therefore, we see the use of a local LLM as pivotal for real-world implementations of our system.
With that, we still see a potential for long-running analytical queries, particularly over vast datasets and architectures like Apache Spark, and when a local LLM instance can be installed.
Especially in these cases, it can be worthwhile to include the expected query runtime or the query complexity into the ranking criteria.
The latter is also interesting in regards to the interpretability of the rewritten queries.
Once there is a good understanding of the quality of the rewrites, it can be interesting to investigate situations when the otherwise accurate rewrite leads to substantially differing insights. This could help to identify wrong facts derived from one data source compared to the evidence found in others.

\balance
\bibliographystyle{ACM-Reference-Format}
\bibliography{main}

@inproceedings{DBLP:conf/sigmod/AikenWH92,
  author       = {Alexander Aiken and
                  Jennifer Widom and
                  Joseph M. Hellerstein},
  editor       = {Michael Stonebraker},
  title        = {Behavior of Database Production Rules: Termination, Confluence, and
                  Observable Determinism},
  booktitle    = {Proceedings of the 1992 {ACM} {SIGMOD} International Conference on
                  Management of Data, San Diego, California, USA, June 2-5, 1992},
  pages        = {59--68},
  publisher    = {{ACM} Press},
  year         = {1992},
  url          = {https://doi.org/10.1145/130283.130296},
  doi          = {10.1145/130283.130296},
  bibsource    = {dblp computer science bibliography, https://dblp.org}
}

@book{dbcompletebook,
author = {Garcia-Molina, Hector and Ullman, Jeffrey D. and Widom, Jennifer},
title = {Database Systems: The Complete Book},
year = {2008},
isbn = {9780131873254},
publisher = {Prentice Hall Press},
address = {USA},
edition = {2}
}

@inproceedings{DBLP:conf/vldb/AhmedLWDSZC06,
  author       = {Rafi Ahmed and
                  Allison W. Lee and
                  Andrew Witkowski and
                  Dinesh Das and
                  Hong Su and
                  Mohamed Za{\"{\i}}t and
                  Thierry Cruanes},
  editor       = {Umeshwar Dayal and
                  Kyu{-}Young Whang and
                  David B. Lomet and
                  Gustavo Alonso and
                  Guy M. Lohman and
                  Martin L. Kersten and
                  Sang Kyun Cha and
                  Young{-}Kuk Kim},
  title        = {Cost-Based Query Transformation in Oracle},
  booktitle    = {Proceedings of the 32nd International Conference on Very Large Data
                  Bases, Seoul, Korea, September 12-15, 2006},
  pages        = {1026--1036},
  publisher    = {{ACM}},
  year         = {2006},
  url          = {http://dl.acm.org/citation.cfm?id=1164215},
  bibsource    = {dblp computer science bibliography, https://dblp.org}
}

@inproceedings{DBLP:conf/btw/0001K15,
  author       = {Thomas Neumann and
                  Alfons Kemper},
  title        = {Unnesting Arbitrary Queries},
  booktitle    = {{BTW}},
  series       = {{LNI}},
  volume       = {{P-241}},
  pages        = {383--402},
  publisher    = {{GI}},
  year         = {2015}
}

@inproceedings{DBLP:conf/btw/000125,
  author       = {Thomas Neumann},
  editor       = {Meike Klettke and
                  Ralf Schenkel and
                  Andreas Henrich and
                  Daniela Nicklas and
                  Maximilian E. Sch{\"{u}}le and
                  Klaus Meyer{-}Wegener},
  title        = {Improving Unnesting of Complex Queries},
  booktitle    = {Datenbanksysteme f{\"{u}}r Business, Technologie und Web {(BTW}
                  2025)},
  series       = {{LNI}},
  volume       = {{P-361}},
  pages        = {25--47},
  publisher    = {Gesellschaft f{\"{u}}r Informatik e.V.},
  year         = {2025},
  url          = {https://doi.org/10.18420/BTW2025-01},
  doi          = {10.18420/BTW2025-01},
  bibsource    = {dblp computer science bibliography, https://dblp.org}
}

@inproceedings{DBLP:conf/icde/SeshadriPL96,
  author       = {Praveen Seshadri and
                  Hamid Pirahesh and
                  T. Y. Cliff Leung},
  title        = {Complex Query Decorrelation},
  booktitle    = {{ICDE}},
  pages        = {450--458},
  publisher    = {{IEEE} Computer Society},
  year         = {1996}
}

@inproceedings{DBLP:conf/sigmod/MumickFPR90,
  author       = {Inderpal Singh Mumick and
                  Sheldon J. Finkelstein and
                  Hamid Pirahesh and
                  Raghu Ramakrishnan},
  title        = {Magic is Relevant},
  booktitle    = {{SIGMOD} Conference},
  pages        = {247--258},
  publisher    = {{ACM} Press},
  year         = {1990}
}

@article{DBLP:journals/tods/Kim82,
  author       = {Won Kim},
  title        = {On Optimizing an SQL-like Nested Query},
  journal      = {{ACM} Trans. Database Syst.},
  volume       = {7},
  number       = {3},
  pages        = {443--469},
  year         = {1982}
}

@article{DBLP:journals/cacm/Miller95,
  author       = {George A. Miller},
  title        = {WordNet: {A} Lexical Database for English},
  journal      = {Commun. {ACM}},
  volume       = {38},
  number       = {11},
  pages        = {39--41},
  year         = {1995},
  url          = {https://doi.org/10.1145/219717.219748},
  doi          = {10.1145/219717.219748},
  bibsource    = {dblp computer science bibliography, https://dblp.org}
}

@article{fellbaum1998wordnet,
  title={WordNet: An electronic lexical database},
  author={Fellbaum, Christiane},
  journal={Cambridge, MA: MIT Press},
  volume={2},
  pages={678--686},
  year={1998}
}

@article{DBLP:journals/jasis/DeerwesterDLFH90,
  author       = {Scott C. Deerwester and
                  Susan T. Dumais and
                  Thomas K. Landauer and
                  George W. Furnas and
                  Richard A. Harshman},
  title        = {Indexing by Latent Semantic Analysis},
  journal      = {Journal of the American society for information science},
  volume       = {41},
  number       = {6},
  pages        = {391--407},
  year         = {1990},
  bibsource    = {dblp computer science bibliography, https://dblp.org}
}

@inproceedings{DBLP:conf/cikm/GuoCXZ11,
  author       = {Jiafeng Guo and
                  Xueqi Cheng and
                  Gu Xu and
                  Xiaofei Zhu},
  editor       = {Craig Macdonald and
                  Iadh Ounis and
                  Ian Ruthven},
  title        = {Intent-aware query similarity},
  booktitle    = {Proceedings of the 20th {ACM} Conference on Information and Knowledge
                  Management, {CIKM} 2011, Glasgow, United Kingdom, October 24-28, 2011},
  pages        = {259--268},
  publisher    = {{ACM}},
  year         = {2011},
  url          = {https://doi.org/10.1145/2063576.2063619},
  doi          = {10.1145/2063576.2063619},
  bibsource    = {dblp computer science bibliography, https://dblp.org}
}

@article{DBLP:journals/corr/abs-2403-14441,
  author       = {Leo K{\"{o}}berlein and
                  Dominik Probst and
                  Richard Lenz},
  title        = {Quantifying Semantic Query Similarity for Automated Linear {SQL} Grading:
                  {A} Graph-based Approach},
  journal      = {CoRR},
  volume       = {abs/2403.14441},
  year         = {2024},
  url          = {https://doi.org/10.48550/arXiv.2403.14441},
  doi          = {10.48550/ARXIV.2403.14441},
  eprinttype    = {arXiv},
  eprint       = {2403.14441},
  bibsource    = {dblp computer science bibliography, https://dblp.org}
}

@article{DBLP:journals/dke/BorodinK20,
  author       = {Gregory Borodin and
                  Yaron Kanza},
  title        = {Search-by-example over {SQL} repositories using structural and intent-driven
                  similarity},
  journal      = {Data Knowl. Eng.},
  volume       = {128},
  pages        = {101811},
  year         = {2020},
  url          = {https://doi.org/10.1016/j.datak.2020.101811},
  doi          = {10.1016/J.DATAK.2020.101811},
  bibsource    = {dblp computer science bibliography, https://dblp.org}
}

@article{DBLP:journals/tkde/KulLXCKU18,
  author       = {G{\"{o}}khan Kul and
                  Duc Thanh Anh Luong and
                  Ting Xie and
                  Varun Chandola and
                  Oliver Kennedy and
                  Shambhu J. Upadhyaya},
  title        = {Similarity Metrics for {SQL} Query Clustering},
  journal      = {{IEEE} Trans. Knowl. Data Eng.},
  volume       = {30},
  number       = {12},
  pages        = {2408--2420},
  year         = {2018},
  url          = {https://doi.org/10.1109/TKDE.2018.2831214},
  doi          = {10.1109/TKDE.2018.2831214},
  bibsource    = {dblp computer science bibliography, https://dblp.org}
}

@article{DBLP:journals/tkde/ArzamasovaBGSS20,
  author       = {Natalia Arzamasova and
                  Klemens B{\"{o}}hm and
                  Bertrand Goldman and
                  Christian Saaler and
                  Martin Sch{\"{a}}ler},
  title        = {On the Usefulness of SQL-Query-Similarity Measures to Find User Interests},
  journal      = {{IEEE} Trans. Knowl. Data Eng.},
  volume       = {32},
  number       = {10},
  pages        = {1982--1999},
  year         = {2020},
  url          = {https://doi.org/10.1109/TKDE.2019.2913381},
  doi          = {10.1109/TKDE.2019.2913381},
  bibsource    = {dblp computer science bibliography, https://dblp.org}
}

@article{DBLP:journals/corr/abs-2403-09060,
  author       = {Jie Liu and
                  Barzan Mozafari},
  title        = {Query Rewriting via Large Language Models},
  journal      = {CoRR},
  volume       = {abs/2403.09060},
  year         = {2024},
  url          = {https://doi.org/10.48550/arXiv.2403.09060},
  doi          = {10.48550/ARXIV.2403.09060},
  eprinttype    = {arXiv},
  eprint       = {2403.09060},
  bibsource    = {dblp computer science bibliography, https://dblp.org}
}

@inproceedings{DBLP:conf/nips/PourrezaR23,
  author       = {Mohammadreza Pourreza and
                  Davood Rafiei},
  title        = {{DIN-SQL:} Decomposed In-Context Learning of Text-to-SQL with Self-Correction},
  booktitle    = {Advances in Neural Information Processing Systems 36: Annual Conference
                  on Neural Information Processing Systems 2023 (NeurIPS)},
  pages = {36339--36348},
  publisher = {Curran Associates, Inc.},
  year         = {2023},
  bibsource    = {dblp computer science bibliography, https://dblp.org}
}

@inproceedings{DBLP:conf/sigir/CarbonellG98,
  author       = {Jaime G. Carbonell and
                  Jade Goldstein},
  title        = {The Use of MMR, Diversity-Based Reranking for Reordering Documents
                  and Producing Summaries},
  booktitle    = {21st Annual International {ACM} {SIGIR}
                  Conference on Research and Development in Information Retrieval},
  pages        = {335--336},
  publisher    = {{ACM}},
  year         = {1998},
  url          = {https://doi.org/10.1145/290941.291025},
  doi          = {10.1145/290941.291025},
  bibsource    = {dblp computer science bibliography, https://dblp.org}
}

@article{DBLP:journals/tkde/Ordonez10,
  author       = {Carlos Ordonez},
  title        = {Optimization of Linear Recursive Queries in {SQL}},
  journal      = {{IEEE} Trans. Knowl. Data Eng.},
  volume       = {22},
  number       = {2},
  pages        = {264--277},
  year         = {2010},
  url          = {https://doi.org/10.1109/TKDE.2009.83},
  doi          = {10.1109/TKDE.2009.83},
  bibsource    = {dblp computer science bibliography, https://dblp.org}
}

@inproceedings{DBLP:conf/sigmod/GanskiW87,
  author       = {Richard A. Ganski and
                  Harry K. T. Wong},
  editor       = {Umeshwar Dayal and
                  Irving L. Traiger},
  title        = {Optimization of Nested {SQL} Queries Revisited},
  booktitle    = {Proceedings of the Association for Computing Machinery Special Interest
                  Group on Management of Data 1987 Annual Conference, San Francisco,
                  CA, USA, May 27-29, 1987},
  pages        = {23--33},
  publisher    = {{ACM} Press},
  year         = {1987},
  url          = {https://doi.org/10.1145/38713.38723},
  doi          = {10.1145/38713.38723},
  bibsource    = {dblp computer science bibliography, https://dblp.org}
}

@article{DBLP:journals/pvldb/Zhou0WLSZ23,
  author       = {Xuanhe Zhou and
                  Guoliang Li and
                  Jianming Wu and
                  Jiesi Liu and
                  Zhaoyan Sun and
                  Xinning Zhang},
  title        = {A Learned Query Rewrite System},
  journal      = {Proc. {VLDB} Endow.},
  volume       = {16},
  number       = {12},
  pages        = {4110--4113},
  year         = {2023},
  url          = {https://www.vldb.org/pvldb/vol16/p4110-li.pdf},
  doi          = {10.14778/3611540.3611633},
  bibsource    = {dblp computer science bibliography, https://dblp.org}
}

@article{DBLP:journals/pvldb/DongLZYMW23,
  author       = {Rui Dong and
                  Jie Liu and
                  Yuxuan Zhu and
                  Cong Yan and
                  Barzan Mozafari and
                  Xinyu Wang},
  title        = {SlabCity: Whole-Query Optimization using Program Synthesis},
  journal      = {Proc. {VLDB} Endow.},
  volume       = {16},
  number       = {11},
  pages        = {3151--3164},
  year         = {2023},
  url          = {https://www.vldb.org/pvldb/vol16/p3151-dong.pdf},
  doi          = {10.14778/3611479.3611515},
  bibsource    = {dblp computer science bibliography, https://dblp.org}
}

@article{DBLP:journals/corr/abs-2310-09716,
  author       = {Fanghua Ye and
                  Meng Fang and
                  Shenghui Li and
                  Emine Yilmaz},
  title        = {Enhancing Conversational Search: Large Language Model-Aided Informative
                  Query Rewriting},
  journal      = {CoRR},
  volume       = {abs/2310.09716},
  year         = {2023},
  url          = {https://doi.org/10.48550/arXiv.2310.09716},
  doi          = {10.48550/ARXIV.2310.09716},
  eprinttype    = {arXiv},
  eprint       = {2310.09716},
  bibsource    = {dblp computer science bibliography, https://dblp.org}
}

@article{DBLP:journals/pvldb/BlunschiJKMS12,
  author       = {Lukas Blunschi and
                  Claudio Jossen and
                  Donald Kossmann and
                  Magdalini Mori and
                  Kurt Stockinger},
  title        = {{SODA:} Generating {SQL} for Business Users},
  journal      = {Proc. {VLDB} Endow.},
  volume       = {5},
  number       = {10},
  pages        = {932--943},
  year         = {2012},
  url          = {http://vldb.org/pvldb/vol5/p932\_lukasblunschi\_vldb2012.pdf},
  doi          = {10.14778/2336664.2336667},
  bibsource    = {dblp computer science bibliography, https://dblp.org}
}

@article{DBLP:journals/corr/abs-1809-08887,
  author       = {Tao Yu and
                  Rui Zhang and
                  Kai Yang and
                  Michihiro Yasunaga and
                  Dongxu Wang and
                  Zifan Li and
                  James Ma and
                  Irene Li and
                  Qingning Yao and
                  Shanelle Roman and
                  Zilin Zhang and
                  Dragomir R. Radev},
  title        = {Spider: {A} Large-Scale Human-Labeled Dataset for Complex and Cross-Domain
                  Semantic Parsing and Text-to-SQL Task},
  journal      = {CoRR},
  volume       = {abs/1809.08887},
  year         = {2018},
  url          = {http://arxiv.org/abs/1809.08887},
  eprinttype    = {arXiv},
  eprint       = {1809.08887},
  bibsource    = {dblp computer science bibliography, https://dblp.org}
}

@inproceedings{levenshtein1966binary,
  title={Binary codes capable of correcting deletions, insertions, and reversals},
  author={Levenshtein, Vladimir I and others},
  booktitle={Soviet physics doklady},
  volume={10},
  issue={8},
  pages={707--710},
  year={1966},
  organization={Soviet Union}
}

@misc{honnibal2020spacy,
  title={spaCy: Industrial-strength natural language processing in python},
  author={Honnibal, Matthew and Montani, Ines and Van Landeghem, Sofie and Boyd, Adriane and others},
  year={2020},
  publisher={Zenodo, Honolulu, HI, USA},
  howpublished = {\url{https://spacy.io/}},
  note = {accessed on 04.02.2025, Copyright 2016-2025}
}

@misc{spacy/models,
  author = {spaCy},
  title = {Englisch spaCy Models},
  howpublished = {\url{https://spacy.io/models/en}},
  note = {accessed on 03.02.2025, Copyright 2016-2025},
  year = {2025}
}

@misc{spacy/models/encore,
  author = {spaCy},
  title = {Large Englisch spaCy Models},
  howpublished = {\url{https://spacy.io/models/en/\#en\_core\_web\_lg}},
  note = {accessed on 03.02.2025, Copyright 2016-2025},
  year = {2025}
}

@misc{openai,
  author = {OpenAI},
  title = {OpenAI Overview},
  howpublished = {\url{https://openai.com/}},
  note = {accessed on 14.02.2025, Copyright 2015-2025},
  year = {2025}
}

@misc{openai/models,
  author = {OpenAI},
  title = {OpenAI Models},
  howpublished = {\url{https://platform.openai.com/docs/models}},
  note = {accessed on 14.02.2025},
  year = {2025}
}

@misc{spider/dataset,
   author  = {Dragomir R. Radev},
  title = {Spider Website},
  howpublished = {\url{https://yale-lily.github.io/spider}},
  note = {accessed on 19.02.2025, data downloaded in November 2024},
  year = {2024}
}

@misc{python/difflib,
   author  = {Python Software Foundation},
  title = {Python Difflib Library},
  howpublished = {\url{https://docs.python.org/3/library/difflib.html}},
  note = {accessed on 04.02.2025, Copyright 2001-2025},
  year = {2025}
}

@misc{openai/models/gpt4,
  author = {OpenAI},
  title = {OpenAI Models GPT-4o},
  howpublished = {\url{https://platform.openai.com/docs/models/gpt-4o\#gpt-4o}},
  note = {accessed on 04.02.2025},
  year = {2025}
}

\end{document}